\documentclass[letterpaper, 10 pt, conference]{ieeeconf}  

\IEEEoverridecommandlockouts                              
\usepackage{template}
\usepackage{algpseudocode}

\usepackage{booktabs}
\usepackage{xcolor}
\usepackage{colortbl}

\algdef{SE}[SUBALG]{Indent}{EndIndent}{}{\algorithmicend\ }%

\algtext*{Indent}
\algtext*{EndIndent}

\numberwithin{paperthm}{subsection}
\usepackage[caption=false, font=footnotesize]{subfig}

\definecolor{Gray}{gray}{0.95}
\definecolor{LightCyan}{rgb}{0.88,1,1}
\newcolumntype{a}{>{\columncolor{Gray}}r}
\newcolumntype{b}{>{\columncolor{white}}r}
\usepackage[backend=bibtex,style=ieee,maxbibnames=6,maxcitenames=2,mincitenames=1,natbib=true]{biblatex}

\title{\LARGE \bf
Multi-Agent Reachability Calibration with Conformal Prediction
}

\author{Anish Muthali$^{1,*}$, Haotian Shen$^{1,*}$, Sampada Deglurkar$^{1}$, Michael H. Lim$^{1}$, \\ Rebecca Roelofs$^{2}$, Aleksandra Faust$^{2}$, Claire Tomlin$^{1}$
\thanks{*Denotes equal contribution.}%
\thanks{$^{1}$Hybrid Systems Lab, EECS, UC Berkeley, USA.}%
\thanks{$^{2}$Google Research, USA.}%
\thanks{Correspondence to {\tt\small anishmuthali@berkeley.edu}.}%
\thanks{Appendices are available in \cite{muthali2023multi}.}
}

\begin{document}

\maketitle
\thispagestyle{empty}
\pagestyle{empty}

\begin{abstract}
We investigate methods to provide safety assurances for autonomous agents that incorporate learning-based predictions of other, uncontrolled agents' behavior into their own trajectory planning. Given a learning-based forecasting model that predicts agents' trajectories, we introduce a method for providing probabilistic assurances on the model's prediction error with calibrated confidence intervals. Through quantile regression, conformal prediction, and reachability analysis, our method generates probabilistically safe and dynamically feasible prediction sets.  We showcase their utility in certifying the safety of planning algorithms, both in simulations using actual autonomous driving data and in an experiment with Boeing vehicles.
\end{abstract}

\section{Introduction}
In safety-critical situations in which an autonomous agent interacts with humans, it is often necessary to predict human behavior in order for the agent to appropriately react. For example, in self-driving tasks, the autonomous car is responsible for assuring the safety of itself and the other vehicles it encounters. State-of-the-art systems try to achieve this through behavior prediction and motion forecasting models, oftentimes black-box neural networks that do not provide interpretation of their inner workings \cite{salzmann2020trajectron++,varadarajan2022multipath++,gu2021densetnt}. However, these models lack rigorous safety assurances, especially in the presence of data distribution shifts. While lower dimensional models can provide safety assurances, these methods rely on parametric assumptions on, for example, a human's rationality \cite{fridovich2020confidence}. These assumptions can be inadequate for complex, multi-modal, and noisy decision-making scenarios. In this work, we provide safety assurances for state-of-the-art black-box trajectory forecasting methods by quantifying these models' uncertainty.  

\begin{figure}[t]
    \centering
    \vspace{0.0625in}
    \includegraphics[width=0.48\textwidth]{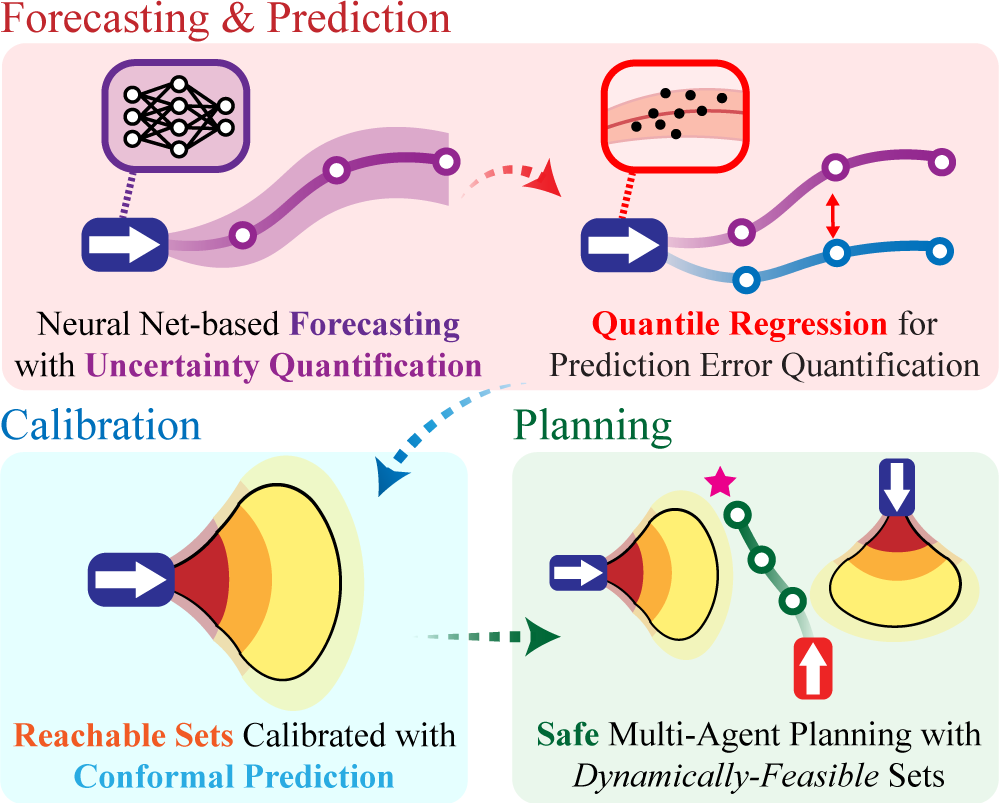}
    \caption{Our method generates dynamically-feasible, probabilistic confidence sets that are derived from a conformal-calibrated quantile regression model.}
    \label{fig:overview}
\end{figure}

However, uncertainty quantification methods alone are insufficient for providing safety assurances. This is because they do not provide guarantees on model behavior and may be unreliable or uninterpretable \cite{kabir2018neural,yao2019quality,charpentier2022disentangling}. Additionally, using prediction model uncertainty for assuredly safe decision-making in downstream planning and control is a challenging problem. 

In our work, we first utilize conformal prediction to \textit{calibrate} measures of uncertainty \cite{vovk2005algorithmic}. Conformal prediction is a statistical tool that uses a heuristic notion of risk to non-parametrically estimate quantiles of risk given a sequence of past observations \cite{angelopoulos2021gentle}. Existing methods that leverage conformal prediction in the context of trajectory forecasting do not explicitly use interpretable metrics of prediction uncertainty \cite{luo2022sample,dixit2022adaptive,tumu2023physics,chen2021reactive,lindemann2023safe}. We propose a method that provides rigorous confidence intervals on model error, a form of probabilistic assurance, given any interpretable heuristics on a trajectory forecasting model's prediction uncertainty. Our approach also allows us to examine the efficacy of various uncertainty quantification heuristics when attempting to predict model error. In addition, we extend our analysis to multi-agent environments to closely reflect real-world assurance cases. We analyze the problem of a single autonomous agent, commonly described as an ``ego agent'', interacting with other, uncontrolled agents.

By producing estimates of model error, we are able to couple statistical assurances with dynamical assurances to allow for safe downstream navigation. In particular, we turn to Hamilton-Jacobi (HJ) reachability analysis \cite{bansal2017hamilton}, which provides guarantees on dynamical systems by means of reachable sets and associated controllers. In HJ reachability, a Hamilton-Jacobi partial differential equation is solved to obtain an optimal value function and controller. The sub-zero level sets of this value function are the reachable sets (possible states of an agent at a given time) and tubes (possible states of an agent \textit{up to} and including a given time).

Our method can be outlined as follows: given a trajectory forecasting model with an associated uncertainty heuristic, we design a quantile regression model that correlates uncertainty with prediction error, creating an approximate confidence interval on the model's prediction. We then apply conformal prediction to calibrate the confidence intervals and provide guarantees on miscoverage rate. We map the calibrated intervals in control action space to sets in state space through reachability analysis, and we demonstrate the utility of these confidence sets in planning tasks. The contributions of this paper include:
\begin{enumerate}
    \item A novel way to interpret trajectory forecasting models' prediction uncertainty and obtain approximate confidence intervals (\Cref{sec:uncertainty});
    \item A technique to calibrate the aforementioned intervals using conformal prediction (\Cref{sec:calibrate});
    \item Dynamically-feasible, probabilistic reachable sets using calibrated intervals (\Cref{sec:prob_reach});
    \item A planning framework that leverages assurances developed in the previous steps (\Cref{sec:safe_plan}).
\end{enumerate}
This paper is organized as follows: \Cref{sec:related_works} discusses related works in conformal prediction and assurances in trajectory forecasting models, \Cref{sec:assurances} and \Cref{sec:prob_reach_plan} describe the contributions outlined above, and \Cref{sec:results} showcases the safety and performance of our methods compared to baseline methods.

\section{Related Works} \label{sec:related_works}
\subsection{Conformal Prediction}
Conformal prediction \cite{vovk2005algorithmic,angelopoulos2021gentle} is a class of uncertainty quantification methods for constructing prediction sets that satisfy a significance level (false negative rate) requirement. Traditionally, conformal prediction creates empirical histograms of measures of risk, called non-conformity scores, and uses these to estimate prediction intervals. Classical techniques include split conformal prediction, which creates empirical histograms from hold-out sets, and full conformal prediction, which creates empirical histograms using all available data \cite{vovk2005algorithmic}. Inductive conformal prediction, a variant of split conformal prediction, uses a non-conformity score that measures distance between train and test data \cite{boursinos2021assurance}. These methods require that the data are identically distributed and exchangeable (any permutation of data points are identically distributed). Methods such as \cite{tibshirani2019conformal} relax the requirement for identically distributed data, and \cite{xu2021conformal,barber2022conformal} relax the exchangeability requirement. Adaptive Conformal Inference \cite{gibbs2021adaptive} and Rolling Risk Control (RollingRC) \cite{feldman2023achieving} have been proposed to relax all assumptions by further calibrating the significance level to match a desired error rate. We adapt RollingRC to provide safety assurances in any multi-agent scenario.

\subsection{Probabilistic Reachability Frameworks}
In this work, we introduce a method to generate probabilistic reachable sets to account for agents' dynamics. Previous work in this space typically involves randomly generating inputs and observing corresponding outputs of a dynamics model, with some associated guarantees in the sampling process \cite{devonport2021data}. Other methods, much like ours, leverage neural network uncertainty \cite{nakamura2022online}. Specifically, the method of \citeauthor{nakamura2022online} \cite{nakamura2022online} uses Gaussian mixture models (GMMs) output by a trajectory forecasting model to generate parametric confidence intervals on control actions, which are then used as control bounds in reachability calculations. In our work, we attempt to relax assumptions that the control actions follow any parametric distribution by applying non-parametric inference techniques.

\subsection{Safety Assurances in Trajectory Prediction}
Various methods for incorporating uncertainty quantification have been examined for the purposes of providing safety assurances in trajectory prediction problems. Some methods provide probabilistic assurances by inferring parameters of a distribution on an agent's control actions \cite{bajcsy2019scalable,fridovich2020confidence}. Methods such as \cite{luo2022sample} and \cite{dixit2022adaptive} estimate confidence intervals with conformal prediction, implicitly leveraging prediction uncertainty through the non-conformity measure. Specifically, the method of \citeauthor{luo2022sample} \cite{luo2022sample} uses split conformal prediction to create a warning system, alerting drivers of ``dangerous'' situations. These warnings can be transformed into confidence sets, as shown in \cite{dixit2022adaptive}, which            additionally eliminates exchangeability assumptions and incorporates trajectory optimization, much like our approach. Other methods emphasize the design process of the trajectory prediction neural networks, for instance by opting to use ReLU networks \cite{chen2021reactive} or by opting to incorporate conformal prediction in the neural network's loss function \cite{tumu2023physics}. In contrast to our approach, none of these methods consider the dynamic feasibility of confidence sets, and some methods that investigate conformal prediction, such as \cite{chen2021reactive} and \cite{luo2022sample}, assume exchangeability. Our method relaxes these assumptions while providing interpretability in the uncertainty quantification process and dynamic feasibility in confidence sets. 

\section{Assurances from Uncertainty} \label{sec:assurances}
Our approach can be summarized in four primary steps: trajectory forecasting with uncertainty quantification (\Cref{sec:traj_forecast}), leveraging uncertainty to obtain approximate prediction intervals (\Cref{sec:uncertainty}), calibrating approximate prediction intervals (\Cref{sec:calibrate}), and obtaining dynamically feasible prediction sets (\Cref{sec:prob_reach}). We summarize our algorithm in \Cref{sec:full_alg}, and we apply our approach to ego agent planning tasks in \Cref{sec:safe_plan}.

\begin{figure*}[t]
    \vspace{0.0625in}
    \centering
    \subfloat[Output of Trajectron++ in a simple scene with two vehicles.\label{fig:traj_output}]{%
        \includegraphics[width=0.30\textwidth]{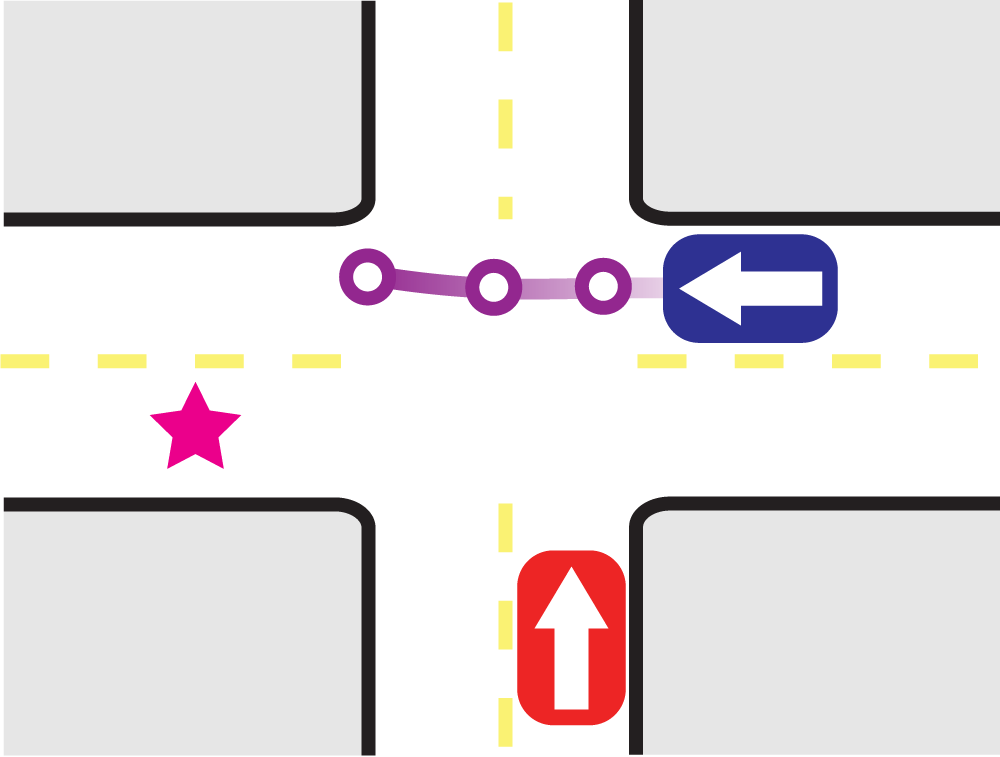}
    }
    \hfill
    \subfloat[Approximate (opaque) and calibrated (translucent) reachable sets.\label{fig:reach}]{%
        \includegraphics[width=0.30\textwidth]{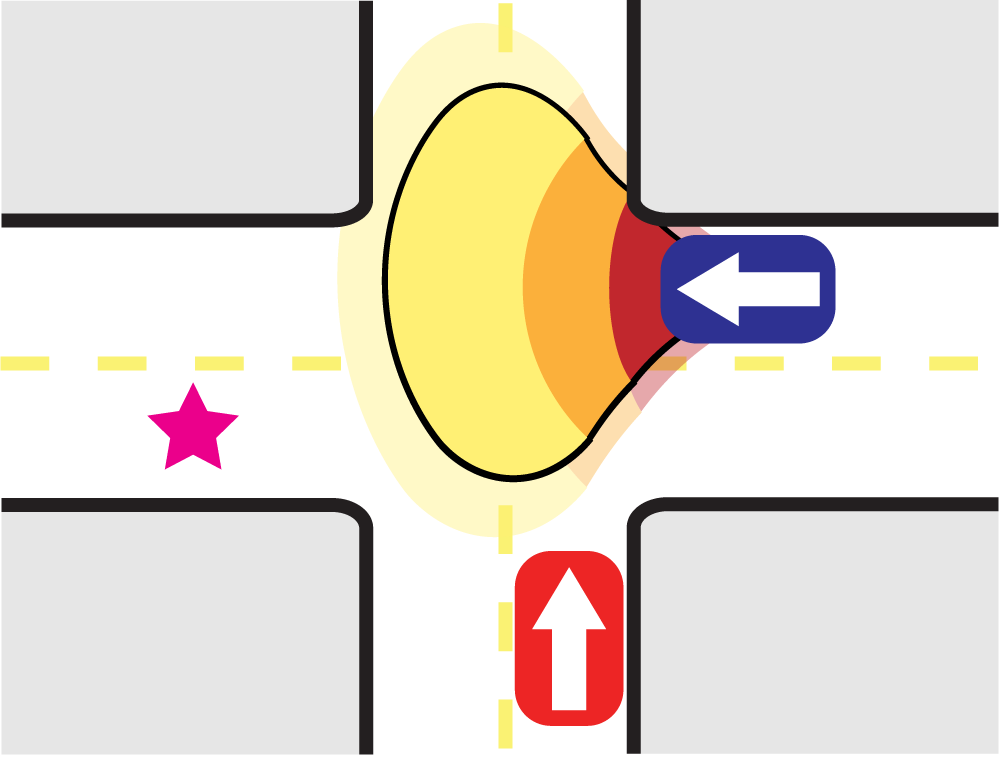}
    }
    \hfill
    \subfloat[Probabilistically-safe plan generated by the reachability-based planner.\label{fig:plan}]{%
        \includegraphics[width=0.30\textwidth]{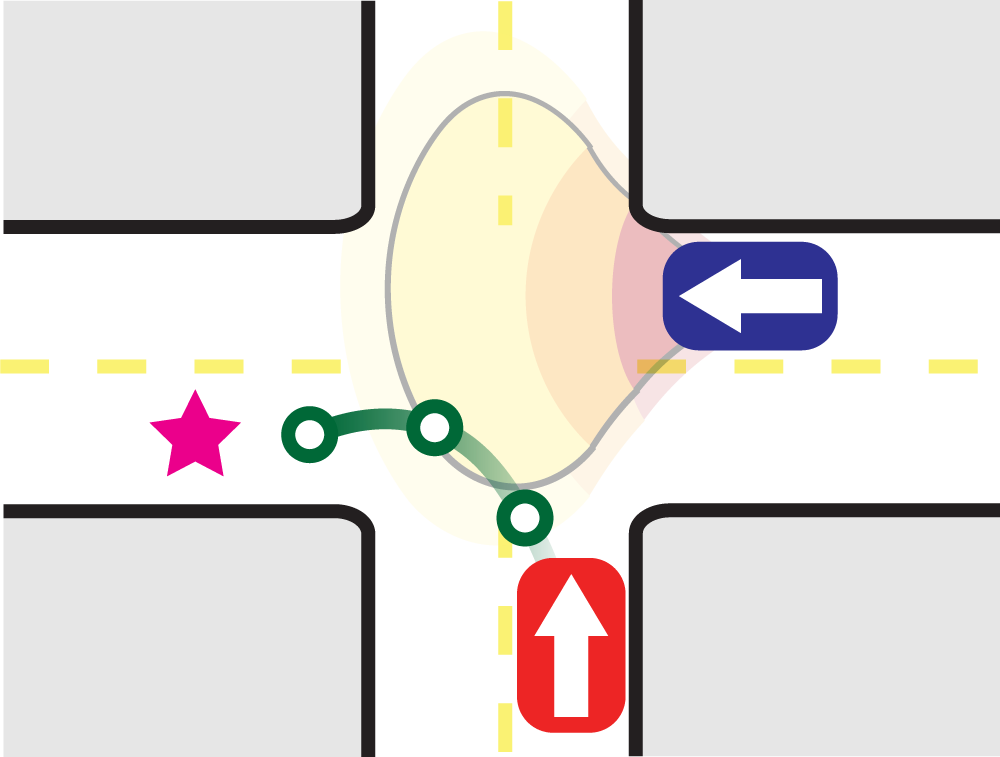}
    }
    \caption{\label{fig:running_ex} Visualization of the running example. The autonomous ego vehicle is shown in red, and the human driver is shown in blue. The ego vehicle aims to navigate to the pink star while avoiding a collision with the human-driven vehicle. Confidence sets for the next three prediction steps are shown. In \cref{fig:reach}, the redder regions represent confidence sets for earlier prediction timesteps, and the translucent regions represent conformal prediction's calibration effect.}
    \vspace{-0.15in}
\end{figure*}

\textit{\textbf{Running Example:} To motivate and illustrate our method, we introduce a simple running example with two vehicles at an intersection, one of them designated as the ``ego'' vehicle. In \Cref{fig:running_ex}, the autonomous ego vehicle, shown in red, aims to safely navigate to the pink-colored star while avoiding the blue vehicle.}

\subsection{Trajectory Forecasting Model} \label{sec:traj_forecast}
We start by assuming access to a known dynamics model for each agent and a trajectory forecasting model capable of predicting an agent's control input. This trajectory forecasting model may maintain the capability to predict control actions for multiple agents at once, while considering interactions between agents. We denote the model as $f_T (\cdot) : \cX \rightarrow \cU$, where $\cX$ is some arbitrary input space and $\cU$ is a space over control actions. The network predicts $\textbf{u}_{t:t+h} \in \cU$, which is a collection of control action vectors indexed by timesteps $t$ through $t+h$, for each of $N$ total agents. Here, $h$ is a fixed prediction horizon. We also assume the existence of an uncertainty measure on the network's outputs, denoted as $\sigma_T(\cdot): \cX \times \cU \rightarrow \bR^d$, with $d$ as the dimension of uncertainty representation. For example, a variance prediction or a variance estimate from inference-time dropout \cite{kabir2018neural} is a valid uncertainty measure. Additionally, some neural network architectures, such as Trajectron++ \cite{salzmann2020trajectron++}, provide alternative uncertainty measures. This network architecture predicts a GMM over possible control actions, leading to understandings of prediction uncertainty such as the variance of the GMM's modes.

\textit{\textbf{Running Example:} Given some sequence of the other vehicle's position history, Trajectron++, our trajectory forecasting model of choice, predicts a GMM in action space, and then integrates the actions to obtain states. We assume that vehicles follow the extended Dubins' car dynamics model. The state of this system is $\mathbf{x} = \mat{x & y & v & \theta}^\top$, and the dynamics are given by $\mathbf{\dot{x}} = \mat{v \cos(\theta) & v \sin(\theta) & u_1 & u_2}^\top$. The variance of the highest-probability Gaussian component, among other features of the GMM, are incorporated into the design of the uncertainty measure.}

\subsection{Estimating Model Error from Uncertainty} \label{sec:uncertainty}
To obtain confidence intervals on a black-box model's outputs, we estimate the neural network's confidence in an online manner, correlating its prediction uncertainty with prediction error. Quantile regression models enable us to map heuristic notions of uncertainty to an \textit{approximate} confidence interval \cite{koenker2005quantile} along each action dimension. We choose a linear model since its simple parametrization allows for fast online updates and interpretability in how it perceives uncertainty. We demonstrate an example of interpretability in \Cref{sec:qr_case_study}. Intuitively, our quantile regression models are approximately ``calibrating'' the network's uncertainty to obtain an estimate of its error.

As we observe new datapoints online, we collect $\mathbf{u}_{t-h:t}$, the last $h$ ground truth control actions prior to timestep $t$. In practice, we estimate control actions by observing the state history of an agent, and then numerically computing derivatives to estimate actions from an assumed dynamics model. We contrast $\mathbf{u}_{t-h:t}$ with the network's previous prediction $h$ timesteps ago, i.e., $\mathbf{\wh{u}}_{t-h:t}$, and define $\mathbf{e}_{t-h:t} \coloneqq \mathbf{u}_{t-h:t} - \mathbf{\wh{u}}_{t-h:t}$ as the prediction error.

Now, suppose we require a $1 - \alpha$ approximate confidence interval on the ground truth control action. We can construct two quantile regression models $\wh{q}_{\frac{\alpha}{2}} : \bR^d \rightarrow \cU$ and $\wh{q}_{1 - \frac{\alpha}{2}} : \bR^d \rightarrow \cU$ where $\wh{q}_\eps$ estimates the $\eps$-quantile on the network's prediction error from timesteps $t$ to $t+h$ for each agent, denoted $\mathbf{\wh{e}}_{\eps}$. For notational convenience, let us denote $\prob{A}[][t]$ as the probability of event $A$ conditioned on information until time $t$. We obtain an approximate $1 - \alpha$ confidence interval as follows:
\begin{align}
    &\prob{\mathbf{\wh{e}}_{\frac{\alpha}{2}} \leq \mathbf{e}_{t:t+h} \leq \mathbf{\wh{e}}_{1-\frac{\alpha}{2}}}[][t]\\
    &=\prob{\mathbf{\wh{e}}_{\frac{\alpha}{2}} \leq \mathbf{u}_{t:t+h} - \mathbf{\wh{u}}_{t:t+h} \leq \mathbf{\wh{e}}_{1-\frac{\alpha}{2}}}[][t] \\
    &=\prob{\mathbf{\wh{u}}_{t:t+h} + \mathbf{\wh{e}}_{\frac{\alpha}{2}} \leq \mathbf{u}_{t:t+h} \leq \mathbf{\wh{u}}_{t:t+h} + \mathbf{\wh{e}}_{1-\frac{\alpha}{2}}}[][t] \\
    &\approx 1 - \alpha.
\end{align}
Thus, our approximate $1 - \alpha$ confidence interval on $\mathbf{u}_{t:t+h}$ is $\wh{\cI}_{t:t+h} = [\mathbf{\wh{u}}_{t:t+h} + \mathbf{\wh{e}}_{\frac{\alpha}{2}}, \mathbf{\wh{u}}_{t:t+h} + \mathbf{\wh{e}}_{1-\frac{\alpha}{2}}]$.

Traditionally, quantile regression models are trained using computationally expensive linear programs \cite{koenker2005quantile}, so instead, we opt for a faster, online gradient descent approach. We define our loss function for the quantile regression model $\wh{q}_\varepsilon$ to be $\cL(y, \wh{y}) = (y - \wh{y}) \varepsilon \mathbf{1}\rc{y \geq \wh{y}} + (\wh{y} - y) (1 - \varepsilon) \mathbf{1}\rc{y < \wh{y}}$ (the ``pinball loss'') \cite{steinwart2011estimating}. We set $y$ to be the true model error, $\mathbf{e}$, and $\wh{y} = \pmb{\beta}^\top \pmb{\sigma}$, where $\pmb{\beta}$ represents the weights of the regression model, and $\pmb{\sigma}$ is the uncertainty measure from the trajectory forecasting model. We update the weights according to $\pmb{\beta} \leftarrow \pmb{\beta} - \zeta \nabla_{\pmb{\beta}} \cL(\mathbf{e}_{t-h:t}, \pmb{\beta})$, with learning rate $\zeta$.

\subsection{Calibrating Approximate Confidence Intervals} \label{sec:calibrate}
Given that the confidence intervals we obtained in the previous section are merely approximate, we aim to calibrate these intervals. To this end, we apply the RollingRC algorithm \cite{feldman2023achieving}, which perfectly adapts to the online requirements of our method. We are motivated to use the RollingRC algorithm compared to other conformal prediction methods due to a desire to remove the data exchangeability assumption, since we allow for sequentially-dependent data and potential distribution shifts. In addition, we would like to train and calibrate the quantile regression models in a sample-efficient manner. RollingRC guarantees that the error rate deviates from $\alpha$ as $\cO\rp{1/T}$, where $T$ is the total number of datapoints provided to the algorithm.

Following the notation from the RollingRC algorithm, we define $\theta_t \in \bR$ as our conformal parameter, and $\vphi(\cdot) : \bR \rightarrow \cU$ as the algorithm's ``stretching function''. Now, we claim that
\begin{equation}
    \label{eq:one_agent_bound}
    \begin{split}
        &\prob{\mathbf{\wh{e}}_{\frac{\alpha}{2}} - \vphi\rp{\pmb{\theta}} \leq \mathbf{e}_{t:t+h} \leq \mathbf{\wh{e}}_{1-\frac{\alpha}{2}} + \vphi\rp{\pmb{\theta}}}[][t] \\
        &\quad \geq 1 - \alpha - \cO\rp{1/t}.
    \end{split}
\end{equation}
Following similar steps as before, we obtain our newly calibrated confidence interval on $\mathbf{u}_{t:t+h}$ as $\cI_{t:t+h} = [\mathbf{\wh{u}}_{t:t+h} + \mathbf{\wh{e}}_{\frac{\alpha}{2}} - \vphi\rp{\pmb{\theta}}, \mathbf{\wh{u}}_{t:t+h} + \mathbf{\wh{e}}_{1-\frac{\alpha}{2}} + \vphi\rp{\pmb{\theta}}]$.

\section{Probabilistic Reachability and Planning} \label{sec:prob_reach_plan}

\subsection{Probabilistic Reachability among Multiple Agents} \label{sec:prob_reach}
In the previous sections, we have designed a method to provide confidence intervals on agents' control actions. However, for some downstream tasks, such as safe planning, confidence sets in spatial dimensions are more desirable. Hence, we use HJ reachability to obtain spatial sets, in the form of forward reachable tubes, on each agent's location given its dynamics and the probabilistic bound on control \cite{schmerling2023}. This procedure asserts that an agent's location will be contained in the produced reachable tube with probability $1 - \alpha$.

Suppose we wish to upper bound the probability that the ego vehicle collides with any agent. Let $\mathbf{x}_t^{(i)}$ be the location of non-ego agent $i$ at timestep $t$ and $\cS[t]^{(i)}$ be the corresponding agent's forward reachable tube, as computed by our algorithm. We define miscoverage rate as the proportion of instances in which the ground truth position of any agent $i$ at time $t$ is outside $\cS[t]^{(i)}$. We aim to obtain an upper bound on miscoverage rate, such that the ego agent can navigate in regions outside of $\cS[t]^{(i)}$ for all $i \in \{1, \ldots, N\}$ and guarantee that the probability of collision is at most $\gamma$, a pre-specified parameter. Consequently, we set the confidence interval significance level $\alpha$ according to our desired total miscoverage rate $\gamma$ and number of agents $N$.
\begin{psettheorem}[Significance Level Correction]
\label{thm:bonferroni}
Suppose that we wish to have a total miscoverage rate of $\gamma$, where total miscoverage rate is an upper bound on the probability that \textit{any} human agent is miscovered:
\begin{equation}
    \label{eq:prob_union}
    \prob{\bigcup_{i = 1}^N \rc{\mathbf{x}_t^{(i)} \not \in \cS[t]^{(i)}}}[][t] \leq \gamma.
\end{equation}
We claim that the following $\alpha$ achieves an (asymptotic) total miscoverage rate of $\gamma$ for $N$ human agents:
\begin{equation}
    \alpha = 1 - \rp{1 - \gamma}^{\frac{1}{N}}.
\end{equation}
\end{psettheorem}
The proof of \Cref{thm:bonferroni} is available in Appendix A, which uses the fact that the $N$ agents act independently conditioned on past information \cite{muthali2023multi}.
Since $\alpha$ must be a fixed quantity in our algorithm, we must also fix $N$. Hence, we fix our algorithm to only consider the $N$ agents closest to the ego vehicle.

\begin{figure}[t]
    \vspace{-0.0625in}
    \begin{algorithm}[H]
        \footnotesize
        \begin{algorithmic}[1]
            \Procedure{GenerateSets}{$\pmb{\theta}$, $\mathbf{\wh{u}}_{t:t+h}$, $\pmb{\sigma}$}
                \State $\mathbf{\wh{e}}_{\frac{\alpha}{2}} \leftarrow \wh{q}_{\frac{\alpha}{2}} \rp{\pmb{\sigma}}$ \Comment{Obtain lower $\frac{\alpha}{2}$ quantile}
                \State $\mathbf{\wh{e}}_{1-\frac{\alpha}{2}} \leftarrow \wh{q}_{1-\frac{\alpha}{2}} \rp{\pmb{\sigma}}$ \Comment{Obtain upper $\frac{\alpha}{2}$ quantile}
                \State $\cI_{t:t+h} \leftarrow \Big[\mathbf{\wh{u}}_{t:t+h} + \mathbf{\wh{e}}_{\frac{\alpha}{2}} - \vphi\rp{\pmb{\theta}},$
                \Indent
                    \Indent
                        $\mathbf{\wh{u}}_{t:t+h} + \mathbf{\wh{e}}_{1-\frac{\alpha}{2}} + \vphi\rp{\pmb{\theta}}\Big]$
                    \EndIndent
                \EndIndent
                \State $\cS \leftarrow \rs{}$
                \For{$t' \in \{t, t+\Delta t, \ldots, t+h\}$}
                    \State $\cS[t'] \leftarrow$ \Call{HJReachability}{$\cI_{t'}$}
                \EndFor
                \State \Return $\cS, \cI_{t:t+h}$
            \EndProcedure
        
            \Procedure{Update}{$\pmb{\theta}$, $\mathbf{u}_{t-h:t}$, $\cI_{t-h:t}$}
                \For{$t' \in \{t, t+\Delta t, \ldots, t+h\}$}
                    \State $\pmb{\theta}\rc{t'} \leftarrow \pmb{\theta}\rc{t'} + \xi \rp{\mathbf{1}\rc{\mathbf{u}_{t'-h} \not \in \cI_{t'-h}} - \alpha}$
                \EndFor
                \State \Call{GradientDescent}{$\wh{q}_{\frac{\alpha}{2}}$, $\mathbf{u}_{t-h:t}$}
                \State \Call{GradientDescent}{$\wh{q}_{1-\frac{\alpha}{2}}$, $\mathbf{u}_{t-h:t}$}
                \State \Return $\wh{q}_{\frac{\alpha}{2}}$, $\wh{q}_{1-\frac{\alpha}{2}}$, $\pmb{\theta}$
            \EndProcedure
    
            \Procedure{Main}{$\gamma$, $N$}
                \State $\alpha \leftarrow 1 - \rp{1 - \gamma}^{\frac{1}{N}}$
                \State $\pmb{\theta}\rc{t,t+\Delta t, \ldots, t+h} \leftarrow 0$
                \State $\wh{q}_{\frac{\alpha}{2}}, \wh{q}_{1-\frac{\alpha}{2}} \leftarrow$ \Call{InitializeRandomWeights}{\,}
                \State $\bI \leftarrow \{\}$
                \State $t \leftarrow 0$
                \While{true}
                    \State $\mathbf{\wh{u}}_{t:t+h}, \pmb{\sigma} \leftarrow f_T\rp{\cdot}, \sigma_T\rp{\cdot}$ \Comment{Get trajectory predictions and uncertainty from model}
                    \State $\cS, \cI_{t:t+h} \leftarrow$ \Call{GenerateSets}{$\pmb{\theta}$, $\mathbf{\wh{u}}_{t:t+h}$, $\pmb{\sigma}$}
                    \State $\bI \leftarrow \bI \cup \cI_{t:t+h}$
                    \If{$t \geq h$}
                        \State $\mathbf{u}_{t-h:t} \leftarrow$ \Call{ObserveHistory}{\,}
                        \State $\cI_{t-h:t} \leftarrow \bI\rs{t-h:t}$
                        \State $\wh{q}_{\frac{\alpha}{2}}$, $\wh{q}_{1-\frac{\alpha}{2}}$, $\pmb{\theta} \leftarrow$ \Call{Update}{$\pmb{\theta}$, $\mathbf{u}_{t-h:t}$, $\cI_{t-h:t}$}
                    \EndIf
                    \State $t \leftarrow t + \Delta t$
                \EndWhile
            \EndProcedure
        \end{algorithmic}
        \caption{\label{alg:full_alg} Conformal Reachability Calibration.}
    \end{algorithm}
\end{figure}

\textit{\textbf{Running Example:} Suppose we want a 95\% probability safety assurance. Since there is only one other vehicle, we get $\alpha = 0.05$ from Theorem \ref{thm:bonferroni}. Given the previous predictions of the blue agent's trajectory, we generate uncalibrated, time-indexed intervals on ranges of possible control actions, denoted $\wh{\cI}_t, \wh{\cI}_{t+\Delta t}, \wh{\cI}_{t+ 2\Delta t}$. We calibrate these using conformal prediction to obtain $\cI_t, \cI_{t+\Delta t}, \cI_{t + 2\Delta t}$. As we explain in the next subsection, HJ reachability allows us to take any sequence of intervals on control actions and generate a time-indexed set of states. In \Cref{fig:reach}, we distinguish the effects of quantile regression and RollingRC's calibration.}

\subsection{Full Algorithm} \label{sec:full_alg}
In \Cref{alg:full_alg}, we demonstrate the final algorithm to generate probabilistic reachable sets. The \textsc{HJReachability} function generates reachable sets given a probabilistic range of control actions, $\cI_{t:t+\Delta t}$. Since the range can differ over time (e.g., $\cI_t \neq \cI_{t+\Delta t}$ necessarily), we iteratively compute time-indexed forward reachable tubes by computing the forward reachable tube over $[t, t + \Delta t]$ and using the reachable \textit{set} at $t + \Delta t$ as the initial condition to compute the reachable tube over $[t + \Delta t, t + 2 \Delta t]$. We also utilize a \textsc{GradientDescent} function that updates the weights of the quantile regression models as described in \Cref{sec:uncertainty}. In \Cref{alg:full_alg}, $\xi$ is the ``learning rate'' associated with the RollingRC algorithm.

\subsection{Safe Planning Framework} \label{sec:safe_plan}
Given the time-indexed sets $\cS[t] \subseteq \cS[t+\Delta t] \subseteq \cdots \subseteq \cS[t+h]$, we desire that the autonomous agent's location at time $t'$ is outside $\cS[t + k \Delta t]$, where $t + (k-1) \Delta t \leq t' \leq t + k \Delta t$. We can plan by treating each agent's time-indexed forward reachable tube as a dynamic obstacle that grows with time. The obstacle-aware planning requirement motivates the application of a forward reach-avoid tube for the ego agent \cite{fisac2015reach,bansal2017hamilton}. We use this to derive an optimal control trajectory by selecting the Hamiltonian-maximizing control trajectory to a desired final state within the forward reach-avoid tube. In practice, this trajectory can involve bang-bang control, so one can track it using a tracker with a provable tracking error bound, such as a constrained iterative linear quadratic regulator. The planner's output is visualized in Appendix D.

\textit{\textbf{Running Example:} From the previous section, we obtained $\cS[t], \cS[t + \Delta t], \cS[t + 2 \Delta t]$ as a probabilistic occupancy region on the location of the other vehicle. Now, we can use the time-varying avoidance regions to plan a safe path to the goal in \Cref{fig:plan}. Notice that the planner allows the ego agent to traverse in the yellow-colored region: it is aware that the ego vehicle would not violate the safety assurance as it can leave the yellow region by the time the other agent would enter it.}

\section{Results} \label{sec:results}
\begin{table*}[ht]
  \vspace{0.0625in}
  \centering
  \caption{Coverage Rates and Set Sizes for $1-\gamma=0.95$.}
  \label{tab:results}
  \begin{tabular}{lababab}
    \toprule
     & \multicolumn{6}{c}{\textbf{Coverage Rates for Prediction Step}}\\
    \cmidrule(l{3pt}r{2pt}){2-7} 
    \textbf{Methods} & 1st (.5s) & 2nd (1s) & 3rd (1.5s) & 4th (2s) & 5th (2.5s) & 6th (3s)\\
    \midrule
    \multicolumn{7}{l}{\textcolor{gray}{\textbf{nuScenes Dataset}}}\\
    \citeauthor{nakamura2022online} & 0.926 \scriptsize{$\pm 0.012$} & 0.854 \scriptsize{$\pm 0.017$} & 0.816 \scriptsize{$\pm 0.023$} & 0.842 \scriptsize{$\pm 0.023$} & 0.868 \scriptsize{$\pm 0.022$} & 0.902 \scriptsize{$\pm 0.020$} \\
    \citeauthor{luo2022sample} & 1.000 \scriptsize{$\pm 0.000$} & 1.000 \scriptsize{$\pm 0.000$} & 1.000 \scriptsize{$\pm 0.000$} & 0.998 \scriptsize{$\pm 0.002$} & 0.989 \scriptsize{$\pm 0.006$} & 0.968 \scriptsize{$\pm 0.012$} \\
    \textbf{Our Method} & 0.964 \scriptsize{$\pm 0.008$} & 0.962 \scriptsize{$\pm$ 0.011} & 0.968 \scriptsize{$\pm 0.010$} & 0.975 \scriptsize{$\pm 0.008$} & 0.981 \scriptsize{$\pm 0.007$}  & 0.985 \scriptsize{$\pm 0.007$}  \\
    \midrule
    \multicolumn{7}{l}{\textcolor{gray}{\textbf{Waymo Dataset}}} \\
    \citeauthor{nakamura2022online} & 0.981 \scriptsize{$\pm 0.007$} & 0.954 \scriptsize{$\pm 0.012$} & 0.938 \scriptsize{$\pm 0.014$} & 0.937 \scriptsize{$\pm 0.015$} & 0.952 \scriptsize{$\pm 0.014$} & 0.955 \scriptsize{$\pm 0.015$} \\
    \citeauthor{luo2022sample} & 1.00 \scriptsize{$\pm 0.00$} & 1.00 \scriptsize{$\pm 0.00$} & 1.00 \scriptsize{$\pm 0.00$} & 0.998 \scriptsize{$\pm 0.002$} & 0.985 \scriptsize{$\pm 0.009$} & 0.955 \scriptsize{$\pm 0.015$} \\
    \textbf{Our Method} & 0.997 \scriptsize{$\pm 0.002$} & 0.986 \scriptsize{$\pm 0.006$} & 0.980 \scriptsize{$\pm 0.008$} & 0.967 \scriptsize{$\pm 0.011$} & 0.965 \scriptsize{$\pm 0.011$}  & 0.965 \scriptsize{$\pm 0.013$}  \\
    \midrule
     & \multicolumn{6}{c}{\textbf{Set Sizes for Prediction Step}}\\
    \cmidrule(l{3pt}r{2pt}){2-7} 
    \multicolumn{7}{l}{\textcolor{gray}{\textbf{nuScenes Dataset}}} \\
    \citeauthor{nakamura2022online} & 57 \scriptsize{$\pm 2$} & 320 \scriptsize{$\pm 12$} & 886 \scriptsize{$\pm 35$} & 2060 \scriptsize{$\pm 85$} & 3259 \scriptsize{$\pm 128$} & 4285 \scriptsize{$\pm 160$} \\
    \citeauthor{luo2022sample} & 425 \scriptsize{$\pm 11$} & 523 \scriptsize{$\pm 16$} & 683 \scriptsize{$\pm 34$} & 1097 \scriptsize{$\pm 109$} & 1426 \scriptsize{$\pm 140$} & 1814 \scriptsize{$\pm 186$} \\
    \textbf{Our Method} & 39 \scriptsize{$\pm 3$} & 157 \scriptsize{$\pm$ 13} & 462 \scriptsize{$\pm 39$} & 1078 \scriptsize{$\pm 88$} & 2150 \scriptsize{$\pm 170$}  & 3713 \scriptsize{$\pm 268$}  \\
    \midrule
    \multicolumn{7}{l}{\textcolor{gray}{\textbf{Waymo Dataset}}}\\
    \citeauthor{nakamura2022online} & 64 \scriptsize{$\pm 7$} & 311 \scriptsize{$\pm 32$} & 951 \scriptsize{$\pm 96$} & 2117 \scriptsize{$\pm 195$} & 3814 \scriptsize{$\pm 328$} & 5892 \scriptsize{$\pm 475$} \\
    \citeauthor{luo2022sample} & 448 \scriptsize{$\pm 8$} & 736 \scriptsize{$\pm 43$} & 1110 \scriptsize{$\pm 94$} & 1568 \scriptsize{$\pm 169$} & 2126 \scriptsize{$\pm 237$} & 2687 \scriptsize{$\pm 301$} \\
    \textbf{Our Method} & 61 \scriptsize{$\pm 6$} & 246 \scriptsize{$\pm 27$} & 655 \scriptsize{$\pm 71$} & 1361 \scriptsize{$\pm 143$} & 2422 \scriptsize{$\pm 240$}  & 3885 \scriptsize{$\pm 365$}  \\
    \bottomrule
  \end{tabular}
  
\end{table*}

We compare the empirical safety and efficiency of our contribution to two baselines, Online Update of Safety Assurances Using Confidence-Based Predictions by \citeauthor{nakamura2022online} \cite{nakamura2022online} and Sample-Efficient Safety Assurances using Conformal Prediction by \citeauthor{luo2022sample} \cite{luo2022sample}. For both baselines, we perform the significance level correction described in \Cref{sec:prob_reach}.

For all benchmarking purposes, we use Trajectron++ trained on the relevant datasets. We follow the same architecture and hyperparameters as \cite{salzmann2020trajectron++} by using 4 seconds (8 steps) of history to predict 3 seconds (6 steps) into the future. This is consistent with the other baselines' approaches. Set sizes are shown in square meters. We use a pre-specified total miscoverage rate of $\gamma = 0.05$, and we generate predictions for the closest $N = 3$ agents, which strikes a balance between the speed of our HJ reachability calculations and the practical safety of the system.

\subsection{nuScenes Dataset Results}
We compare the coverage rate and efficiency of our method against the two baselines on nuScenes self-driving data \cite{nuscenes}. We calculate average coverage rate and average set sizes individually for each forward prediction step $t, t + \Delta t, \ldots, t + h$ on 100 randomly sampled scenes. For each scene, we use the first 13 seconds to calibrate each method and make predictions on the last 5.5 seconds. \Cref{tab:results} shows step coverage and set sizes at all prediction steps. Note that an ideal algorithm maintains a coverage rate over $1 - \gamma$ while providing the smallest prediction sets.

\subsection{Waymo Open Motion Dataset Results}
To demonstrate the planning safety and efficiency of each method, we also perform experiments on the Waymo Open Motion Dataset \cite{Ettinger_2021_ICCV}, coupled with the Nocturne simulator \cite{vinitsky2022nocturne}. This allows us to apply control actions to the ego vehicle while all other agents replay their respective sequences of control actions from the dataset. We use the same planning method discussed in \Cref{sec:safe_plan} for all three methods, since neither of the baselines have associated planners. For each scene, we calibrate using the first 7 seconds and use model-predictive control to plan for the last 3 seconds, where the goal is the final position of the ego vehicle in the ground truth data. We measure three quantities: (1) progress to goal, defined as the ratio of the distance from the final state of the ego vehicle to the goal compared to the distance from the start to the goal, subtracted from 1; (2) collision rate; (3) conservatism of each method compared to the ego vehicle's ground truth trajectory, defined as the ratio of minimum distance between the ego vehicle to other agents at all times as a result of the planner, compared to that of the ground truth. The formulas and computations of these metrics are described in detail in Appendix B \cite{muthali2023multi}. In Appendix C, we additionally demonstrate the impact of the aforementioned theoretical guarantees by providing safety and efficiency metrics in the absence of conformal prediction \cite{muthali2023multi}. We performed the benchmarks on 200 randomly sampled scenes. \Cref{tab:results} depicts coverage rates and set sizes for all prediction timesteps. \Cref{tab:waymo_collision} depicts average collision rate, average progress to goal, and conservatism.

\begin{table}[t]
  \centering
  \scriptsize
  \caption{Waymo Planning Benchmarks}
    \label{tab:waymo_collision}
  \begin{tabular}{laba}
    \toprule
    \textbf{Method} & \textbf{Progress} & \textbf{Collision} & \textbf{Conservatism} \\
    & \textbf{to Goal} & \textbf{Rate} & \\
    \midrule
    \citeauthor{nakamura2022online} & 0.494 \scriptsize{$\pm 0.029$} & \textbf{0.0} & \textbf{1.504} \scriptsize{$\pm 0.068$} \\
    \citeauthor{luo2022sample} & 0.305 \scriptsize{$\pm 0.028$} & 0.005 & 1.626 \scriptsize{$\pm 0.072$} \\
    \textbf{Our Method} & \textbf{0.544} \scriptsize{$\pm 0.028$} & \textbf{0.0} & 1.507 \scriptsize{$\pm 0.068$} \\
    \bottomrule
  \end{tabular}
    
\end{table}

\begin{figure*}[t]
  \vspace{0.0625in}
    \centering
    \subfloat[Sample scenario in which we observe the reachable sets of the three agents closest to the ego vehicle.\label{fig:qr_empty}]{%
        \includegraphics[width=0.30\textwidth]{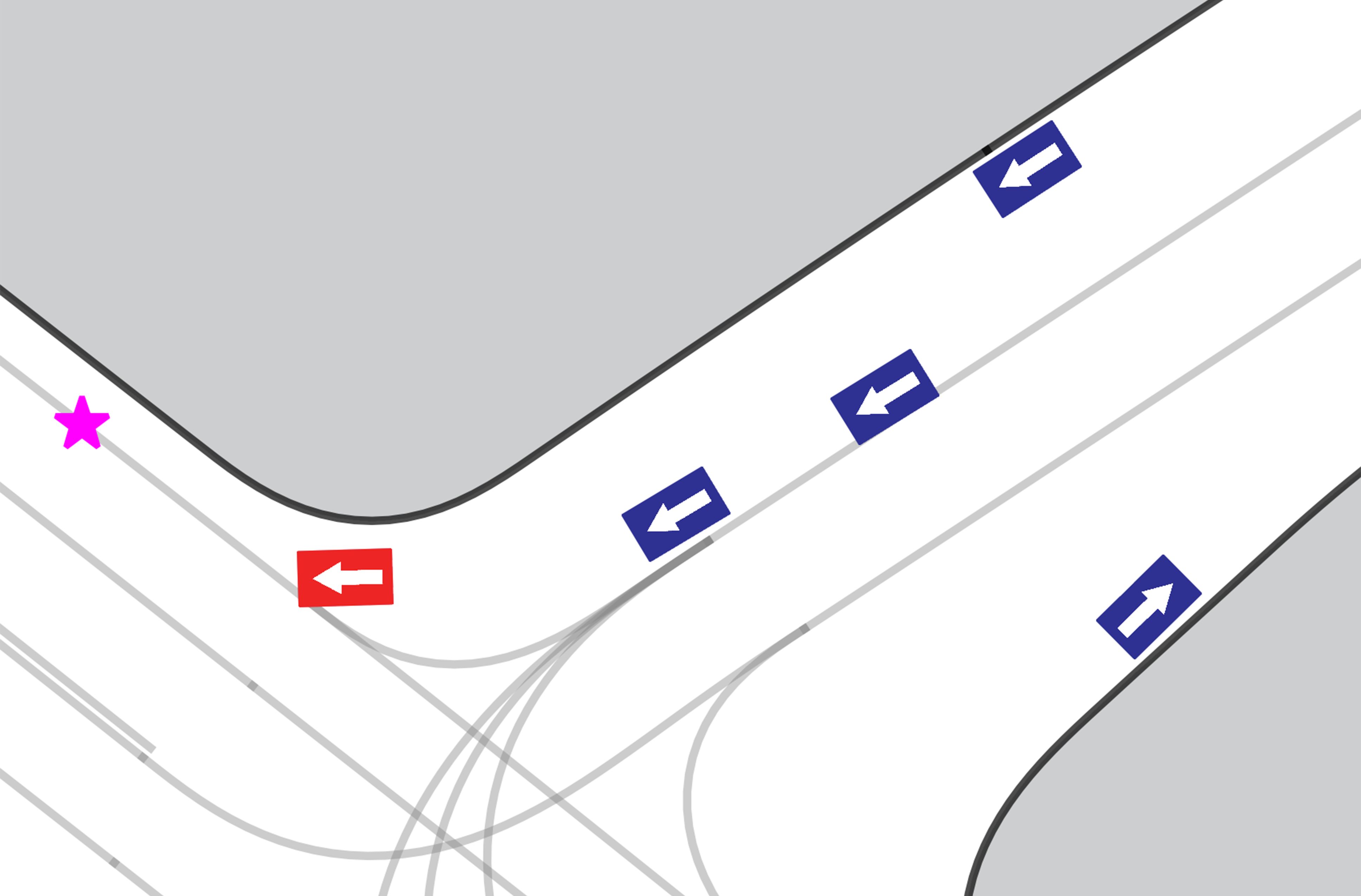}
    }
    \hfill
    \subfloat[Calibrated confidence sets generated by quantile regression \textbf{without} covariance features.\label{fig:qr_no_cov}]{%
        \includegraphics[width=0.30\textwidth]{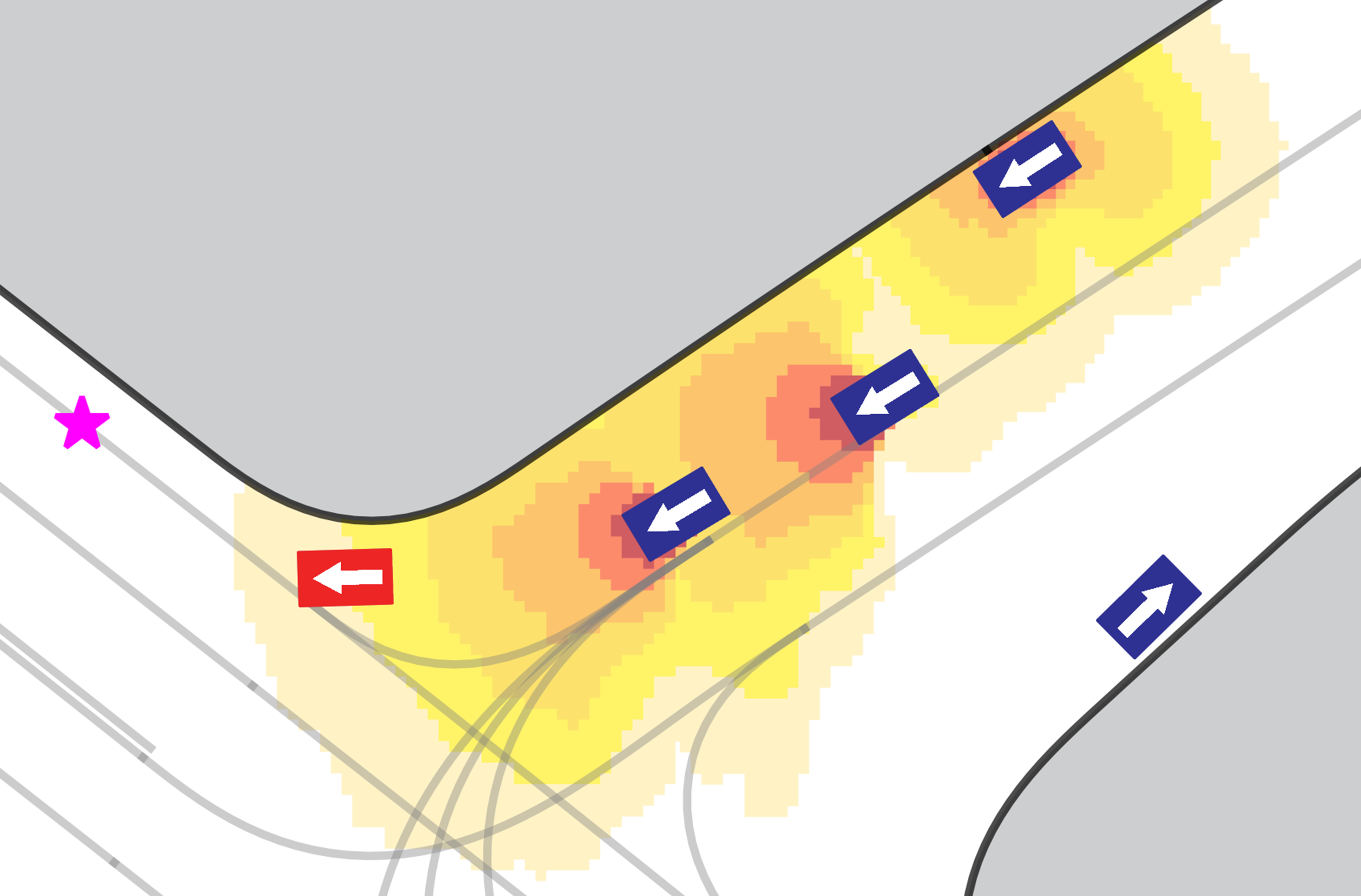}
    }
    \hfill
    \subfloat[Calibrated confidence sets generated by quantile regression \textbf{with} covariance features.\label{fig:qr_all_feat}]{%
        \includegraphics[width=0.30\textwidth]{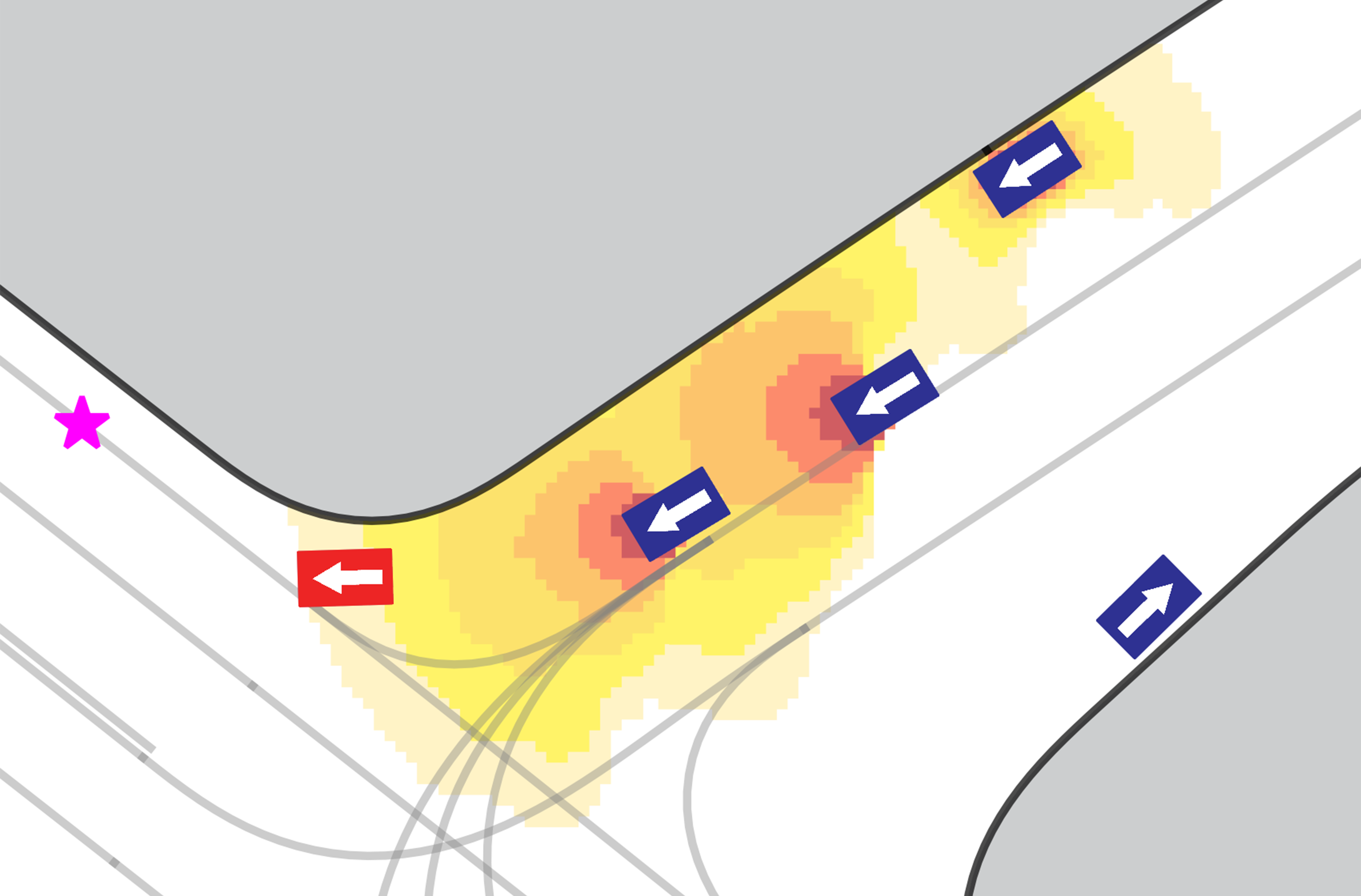}
    }
    \caption{\label{fig:qr_case_study} Case Study of Uncertainty Metrics. We demonstrate a simple example in which the choice of uncertainty measure affects the size of sets, with coverage rate held constant.}
\end{figure*}

\subsection{Discussion of Results}
For the nuScenes dataset, we notice that our method achieves more efficient set sizes for initial prediction steps, while \citeauthor{luo2022sample} achieves more efficient set sizes for later prediction timesteps. Nevertheless, neither of these two methods violates the miscoverage requirement of $\gamma = 0.05$. The method of \citeauthor{nakamura2022online} violates the miscoverage rate, however, supporting the introduction of uncertainty calibration into the algorithm. Hence, calibrating neural network uncertainty is important, not only to provide the desired coverage rate but also to generate efficient prediction sets. 

For the Waymo dataset, we notice a very similar phenomenon with set sizes and coverage rates. 
In the planning benchmarks, our method has the best progress to goal, likely due to the initial-timestep sets being smaller. This is also reflected in the conservatism scores, with reachability-based methods performing the best. The method of \citeauthor{luo2022sample} also encountered one collision scenario in which the produced set was very large and forced the planner to take a sharp avoid action. Thus, we note the importance of initial-timestep sets being small to allow the reachability-based methods to perform better in the planning benchmarks. This allows the ego vehicle to make some progress, whereas a large initial-timestep set would inhibit any progress regardless of the relative size of later timesteps' sets.

\subsection{Case Studies}

\begin{figure}[t]
    \centering
    \vspace{0.0625in}
    \includegraphics[width=0.4\textwidth]{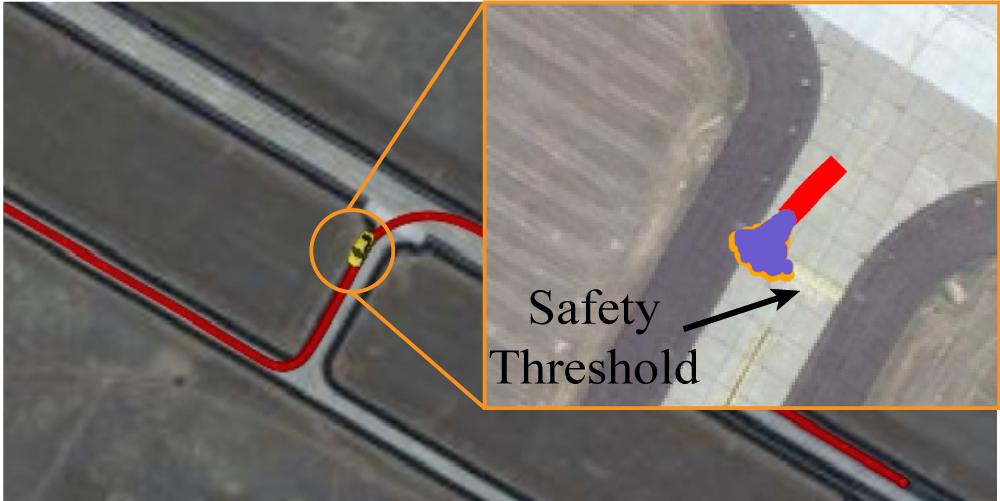}
    \caption{Our algorithm is applied to assure safety in potential runway incursion scenarios. Once the ground vehicle is determined to have crossed a designated safety threshold, the aircraft is cleared to land.}
    \label{fig:boeing}
\end{figure}

\subsubsection{Understanding Uncertainty Measures} \label{sec:qr_case_study}
In this case study, we demonstrate the usefulness of our interpretable quantile regression model when understanding the efficacy of uncertainty metrics. Consider the scene in \Cref{fig:qr_empty}. We choose the uncertainty measure based on properties of the GMM, including the distance between peaks and the (co)variance of the highest-weighted mode. The learned regression model indicates a positive correlation between prediction error and variance of the most-likely GMM mode. In \Cref{fig:qr_no_cov} and \Cref{fig:qr_all_feat}, we can visually discern the benefit of including these features.

Overall, this case study shows the importance of understanding the usefulness of different components of the uncertainty measure. A more useful uncertainty metric can provide more efficient sets, since a more accurate quantile regression model would require less calibration (less ``stretching'' from conformal prediction). Conformal prediction cannot derive confidence intervals conditional on some input, so quantile regression's accuracy is crucial for providing efficient sets.

\subsubsection{Safety in Aerospace Applications}
In this case study, we apply our algorithm to satisfy a real-world safety assurance requirement by demonstrating our algorithm on Boeing vehicles. We consider the case of an aircraft attempting to land on a runway while accounting for potential runway incursions from ground vehicles. We use our algorithm to provide assurances on the motion of a ground vehicle on the runway. Given a fixed landing plan for the plane, we adapt the sets from our algorithm to design a warning system similar to \citeauthor{luo2022sample}'s original algorithm. If a prediction set intersects the runway, a warning is issued. A visualization of this application is shown in \Cref{fig:boeing}. The ground vehicle's state history is shown in red, and its uncalibrated prediction set is shown in purple. The ``stretching'' effect from conformal prediction is shown in orange.

\section{Discussion and Future Work} \label{sec:fut_work}
In this paper, we introduced a non-parametric approach to using interpretable uncertainty measures from black-box models for generating calibrated prediction intervals. We demonstrated an efficient reachability-based approach to generating prediction sets, and we showed the goal-oriented efficiency and safety of our algorithm in planning tasks for an ego agent through simulations and real-world experiments.

For future investigations, we would be interested in seeing the effects of longer planning horizons. Although many state of the art models cannot provide reliable predictions for agent behavior 1 minute into the future, for example, we would like to see the efficiency of our method compared to the other methods discussed. We would also like to generalize our method to arbitrary measures of risk, instead of only coverage rate. For example, one might want greater confidence in the behavior of nearby or fast-moving agents, than for agents that are far away or stationary. Thus, in certain practical scenarios, a heuristic measure of risk may be more appropriate than miscoverage rate. Along these lines, we would also like to explore adapting the confidence interval significance $\alpha$ for different agents depending on their properties with respect to maintaining safety. Finally, we are interested in decreasing conservatism across the pipeline to help ensure that the planning framework can always find a feasible solution to the problem. 

\section{Acknowledgements}
This material is based upon work supported by the DARPA Assured Autonomy Program, the SRC CONIX program, Google-BAIR Commons, the NASA ULI program on Safe Aviation Autonomy, and the National Science Foundation Graduate Research Fellowship Program under Grant Nos. DGE 1752814 and DGE 2146752.
Any opinions, findings, and conclusions or recommendations expressed in this material are those of the authors and do not necessarily reflect the views of any aforementioned organizations. 
We also thank James Paunicka, Blake Edwards, Dragos Margineantu, Douglas Stuart, and the entire team at Boeing for all their help and contributions in demonstrating our method on Boeing's runway incursion dataset. Finally, we thank Marius Wiggert for helping with the implementations in the HJ reachability toolbox.

\printbibliography

@inproceedings{devonport2021data,
  title={Data-driven reachability analysis with Christoffel functions},
  author={Devonport, Alex and Yang, Forest and El Ghaoui, Laurent and Arcak, Murat},
  booktitle={2021 60th IEEE Conference on Decision and Control},
  pages={5067--5072},
  year={2021},
  organization={IEEE}
}

@article{feldman2023achieving,
      title={Achieving Risk Control in Online Learning Settings}, 
      author={Shai Feldman and Liran Ringel and Stephen Bates and Yaniv Romano},
      year={2023},
      journal={arXiv preprint 2205.09095},
}

@article{fridovich2020confidence,
  title={Confidence-aware motion prediction for real-time collision avoidance},
  author={Fridovich-Keil, David and Bajcsy, Andrea and Fisac, Jaime F and Herbert, Sylvia L and Wang, Steven and Dragan, Anca D and Tomlin, Claire J},
  journal={The International Journal of Robotics Research},
  volume={39},
  number={2-3},
  pages={250--265},
  year={2020},
  publisher={SAGE Publications Sage UK: London, England}
}

@article{vinitsky2022nocturne,
  title={Nocturne: a scalable driving benchmark for bringing multi-agent learning one step closer to the real world},
  author={Vinitsky, Eugene and Lichtl{\'e}, Nathan and Yang, Xiaomeng and Amos, Brandon and Foerster, Jakob},
  journal={arXiv preprint 2206.09889},
  year={2022}
}

@article{nakamura2022online,
  title={Online update of safety assurances using confidence-based predictions},
  author={Nakamura, Kensuke and Bansal, Somil},
  journal={arXiv preprint 2210.01199},
  year={2022}
}

@article{dixit2022adaptive,
  title={Adaptive Conformal Prediction for Motion Planning among Dynamic Agents},
  author={Dixit, Anushri and Lindemann, Lars and Wei, Skylar and Cleaveland, Matthew and Pappas, George J and Burdick, Joel W},
  journal={arXiv preprint 2212.00278},
  year={2022}
}

@article{tumu2023physics,
  title={Physics Constrained Motion Prediction with Uncertainty Quantification},
  author={Tumu, Renukanandan and Lindemann, Lars and Nghiem, Truong and Mangharam, Rahul},
  journal={arXiv preprint 2302.01060},
  year={2023}
}

@inproceedings{salzmann2020trajectron++,
  title={Trajectron++: Dynamically-feasible trajectory forecasting with heterogeneous data},
  author={Salzmann, Tim and Ivanovic, Boris and Chakravarty, Punarjay and Pavone, Marco},
  booktitle={Computer Vision--ECCV 2020: 16th European Conference, Glasgow, UK, August 23--28, 2020, Proceedings, Part XVIII 16},
  pages={683--700},
  year={2020},
  organization={Springer}
}

@inproceedings{luo2022sample,
  title={Sample-efficient safety assurances using conformal prediction},
  author={Luo, Rachel and Zhao, Shengjia and Kuck, Jonathan and Ivanovic, Boris and Savarese, Silvio and Schmerling, Edward and Pavone, Marco},
  booktitle={Algorithmic Foundations of Robotics XV: Proceedings of the Fifteenth Workshop on the Algorithmic Foundations of Robotics},
  pages={149--169},
  year={2022},
  organization={Springer}
}

@inproceedings{chen2021reactive,
  title={Reactive motion planning with probabilistic safety guarantees},
  author={Chen, Yuxiao and Rosolia, Ugo and Fan, Chuchu and Ames, Aaron and Murray, Richard},
  booktitle={Conference on Robot Learning},
  pages={1958--1970},
  year={2021},
  organization={PMLR}
}

@inproceedings{bansal2017hamilton,
  title={Hamilton-jacobi reachability: A brief overview and recent advances},
  author={Bansal, Somil and Chen, Mo and Herbert, Sylvia and Tomlin, Claire J},
  booktitle={2017 IEEE 56th Annual Conference on Decision and Control},
  pages={2242--2253},
  year={2017},
  organization={IEEE}
}

@book{koenker2005quantile,
  title={Quantile regression},
  author={Koenker, Roger},
  volume={38},
  year={2005},
  publisher={Cambridge university press}
}

@article{steinwart2011estimating,
  title={Estimating conditional quantiles with the help of the pinball loss},
  author={Steinwart, Ingo and Christmann, Andreas},
  journal={arXiv preprint 1102.2101},
  year={2011}
}

@inproceedings{fisac2015reach,
  title={Reach-avoid problems with time-varying dynamics, targets and constraints},
  author={Fisac, Jaime F and Chen, Mo and Tomlin, Claire J and Sastry, S Shankar},
  booktitle={Proceedings of the 18th International Conference on Hybrid Systems: Computation and Control},
  pages={11--20},
  year={2015}
}

@inproceedings{Ettinger_2021_ICCV,
    author={Ettinger, Scott and Cheng, Shuyang and Caine, Benjamin and Liu, Chenxi and Zhao, Hang and Pradhan, Sabeek and Chai, Yuning and Sapp, Ben and Qi, Charles R. and Zhou, Yin and Yang, Zoey and Chouard, Aur'elien and Sun, Pei and Ngiam, Jiquan and Vasudevan, Vijay and McCauley, Alexander and Shlens, Jonathon and Anguelov, Dragomir},
    title={Large Scale Interactive Motion Forecasting for Autonomous Driving: The Waymo Open Motion Dataset},
    booktitle={International Conference on Computer Vision}, month={October}, year={2021}, pages={9710-9719},
  organization={IEEE}
}

@inproceedings{nuscenes,
  title={{nuScenes}: A multimodal dataset for autonomous driving},
  author={Holger Caesar and Varun Bankiti and Alex H. Lang and Sourabh Vora and 
          Venice Erin Liong and Qiang Xu and Anush Krishnan and Yu Pan and 
          Giancarlo Baldan and Oscar Beijbom}, 
  booktitle={Computer Vision and Pattern Recognition},
  year=2020,
  organization={IEEE}
}

@book{vovk2005algorithmic,
  title={Algorithmic learning in a random world},
  author={Vovk, Vladimir and Gammerman, Alexander and Shafer, Glenn},
  volume={29},
  year={2005},
  publisher={Springer}
}

@article{tibshirani2019conformal,
  title={Conformal prediction under covariate shift},
  author={Tibshirani, Ryan J and Foygel Barber, Rina and Candes, Emmanuel and Ramdas, Aaditya},
  journal={Advances in Neural Information Processing Systems},
  volume={32},
  year={2019}
}

@article{angelopoulos2021gentle,
  title={A gentle introduction to conformal prediction and distribution-free uncertainty quantification},
  author={Angelopoulos, Anastasios N and Bates, Stephen},
  journal={arXiv preprint 2107.07511},
  year={2021}
}

@inproceedings{xu2021conformal,
  title={Conformal prediction interval for dynamic time-series},
  author={Xu, Chen and Xie, Yao},
  booktitle={International Conference on Machine Learning},
  pages={11559--11569},
  year={2021},
  organization={PMLR}
}

@article{barber2022conformal,
  title={Conformal prediction beyond exchangeability},
  author={Barber, Rina Foygel and Candes, Emmanuel J and Ramdas, Aaditya and Tibshirani, Ryan J},
  journal={arXiv preprint 2202.13415},
  year={2022}
}

@article{gibbs2021adaptive,
  title={Adaptive conformal inference under distribution shift},
  author={Gibbs, Isaac and Candes, Emmanuel},
  journal={Advances in Neural Information Processing Systems},
  volume={34},
  pages={1660--1672},
  year={2021}
}

@inproceedings{varadarajan2022multipath++,
  title={Multipath++: Efficient information fusion and trajectory aggregation for behavior prediction},
  author={Varadarajan, Balakrishnan and Hefny, Ahmed and Srivastava, Avikalp and Refaat, Khaled S and Nayakanti, Nigamaa and Cornman, Andre and Chen, Kan and Douillard, Bertrand and Lam, Chi Pang and Anguelov, Dragomir and others},
  booktitle={International Conference on Robotics and Automation},
  pages={7814--7821},
  year={2022},
  organization={IEEE}
}

@inproceedings{gu2021densetnt,
  title={Densetnt: End-to-end trajectory prediction from dense goal sets},
  author={Gu, Junru and Sun, Chen and Zhao, Hang},
  booktitle={International Conference on Computer Vision},
  pages={15303--15312},
  year={2021},
  organization={IEEE}
}

@article{kabir2018neural,
  title={Neural network-based uncertainty quantification: A survey of methodologies and applications},
  author={Kabir, HM Dipu and Khosravi, Abbas and Hosen, Mohammad Anwar and Nahavandi, Saeid},
  journal={IEEE access},
  volume={6},
  pages={36218--36234},
  year={2018},
  publisher={IEEE}
}

@article{yao2019quality,
  title={Quality of uncertainty quantification for Bayesian neural network inference},
  author={Yao, Jiayu and Pan, Weiwei and Ghosh, Soumya and Doshi-Velez, Finale},
  journal={arXiv preprint 1906.09686},
  year={2019}
}

@article{charpentier2022disentangling,
  title={Disentangling epistemic and aleatoric uncertainty in reinforcement learning},
  author={Charpentier, Bertrand and Senanayake, Ransalu and Kochenderfer, Mykel and Gunnemann, Stephan},
  journal={arXiv preprint 2206.01558},
  year={2022}
}

@inproceedings{bajcsy2019scalable,
  title={A scalable framework for real-time multi-robot, multi-human collision avoidance},
  author={Bajcsy, Andrea and Herbert, Sylvia L and Fridovich-Keil, David and Fisac, Jaime F and Deglurkar, Sampada and Dragan, Anca D and Tomlin, Claire J},
  booktitle={2019 International Conference on Robotics and Automation},
  pages={936--943},
  year={2019},
  organization={IEEE}
}

@article{boursinos2021assurance,
  title={Assurance monitoring of learning-enabled cyber-physical systems using inductive conformal prediction based on distance learning},
  author={Boursinos, Dimitrios and Koutsoukos, Xenofon},
  journal={AI EDAM},
  volume={35},
  number={2},
  pages={251--264},
  year={2021},
  publisher={Cambridge University Press}
}

@misc{schmerling2023,
  author = {Schmerling, Edward},
  title = {hj\_reachability},
  year = {2021},
  publisher = {GitHub},
  journal = {GitHub repository},
  howpublished = {\url{https://github.com/StanfordASL/hj_reachability}},
  commit = {2ae67d252004badc0aa62e5d7bf28b1acdaa9a4c}
}

@article{lindemann2023safe,
  title={Safe planning in dynamic environments using conformal prediction},
  author={Lindemann, Lars and Cleaveland, Matthew and Shim, Gihyun and Pappas, George J},
  journal={IEEE Robotics and Automation Letters},
  year={2023},
  publisher={IEEE}
}

@article{muthali2023multi,
  title={Multi-agent reachability calibration with conformal prediction},
  author={Muthali, Anish and Shen, Haotian and Deglurkar, Sampada and Lim, Michael H and Roelofs, Rebecca and Faust, Aleksandra and Tomlin, Claire},
  journal={arXiv preprint arXiv:2304.00432},
  year={2023}
}


\clearpage

\appendix

\setcounter{psettheorem}{0}

\subsection{Proof of Theorem \ref{thm:bonferroni}}\label{app:bonferroniproof}
The theorem is restated for convenience below:
\begin{psettheorem}[Significance Level Correction]
    Suppose that we wish to have a total miscoverage rate of $\gamma$, where total miscoverage rate is an upper bound on the probability that \textit{any} human agent is miscovered:
    \begin{equation}
        \label{eq:coverage_prob}
        \prob{\bigcup_{i = 1}^N \rc{\mathbf{x}_t^{(i)} \not \in \cS[t]^{(i)}}}[][t] \leq \gamma.
    \end{equation}
    We claim that the following $\alpha$ achieves an (asymptotic) total miscoverage rate of $\gamma$ for $N$ human agents:
    \begin{equation}
        \alpha = 1 - \rp{1 - \gamma}^{\frac{1}{N}}.
    \end{equation}
\end{psettheorem}
\begin{proof}
    To obtain this, we assume that each agent reacts to other agents entirely based on past observations. Mathematically, we define a filtration over the probability space of all agents' time-dependent actions $\cF_0 \subset \cF_{\Delta t} \subset \ldots \subset \cF_{t - \Delta t} \subset \cF_t$. We assume that any probability event corresponding to agent $i$'s future behavior, denoted $A_{t'}^{(i)}$, is conditionally independent of any probability event corresponding to agent $j$'s future behavior, denoted $A_{t'}^{(j)}$. Implicitly, we write $t' > t$. Specifically,
    \begin{equation}
        \prob{A_{t'}^{(i)} A_{t'}^{(j)}}[\cF_t] = \prob{A_{t'}^{(i)}}[\cF_t] \prob{A_{t'}^{(j)}}[\cF_t].
    \end{equation}
    Intuitively, no agent makes decisions on unseen observations. Hence, we may rewrite \cref{eq:coverage_prob} as follows using the conditional independence and the corresponding lower bound on coverage rate for each agent:
    \begin{align}
        &\prob{\bigcup_{i = 1}^N \rc{\mathbf{x}_t^{(i)} \not \in \cS[t]^{(i)}}}[\cF_t] \\
        &= 1 - \prob{\bigcap_{i = 1}^N \rc{\mathbf{x}_t^{(i)} \in \cS[t]^{(i)}}}[\cF_t] \\
        &= 1 - \prod_{i = 1}^N \underbrace{\prob{\mathbf{x}_t^{(i)} \in \cS[t]^{(i)}}[\cF_t]}_{\geq 1 - \alpha - \cO\rp{1/T}} \\
        &\leq 1 - \prod_{i = 1}^N \rp{1 - \alpha - \cO\rp{1/T}} \\
        &= 1 - \rp{1 - \alpha - \cO\rp{1/T}}^N \leq \gamma.
    \end{align}
    Since we aim for an asymptotic bound, we take $T \rightarrow \infty$ which allows us to omit the $\cO\rp{1/T}$ term. To obtain the smallest value of $\alpha$ possible, we meet the loose inequality with equality, and find $\alpha$ such that $1 - (1 - \alpha)^N = \gamma$. Solving for $\alpha$ yields the value $\alpha = 1 - (1 - \gamma)^{\frac{1}{N}}$.
\end{proof}

\subsection{Metrics Computation}\label{app:metrics}
\begin{itemize}
    \item \underline{Coverage rate:} Letting previous notation prevail, we determine $k^{th}$ prediction step coverage rate for any given scene by computing
    \begin{equation*}
        \mathbf{1}\rc{\bigcap_{i = 1}^N \rc{\mathbf{x}_{t + k \Delta t}^{(i)} \in \cS[t+k \Delta t]^{(i)}}}    
    \end{equation*}
    per timestep. We average this quantity over timesteps to obtain average coverage rate for the specific scene.

    \item \underline{Conservatism:} Define $\mathbf{x}_t^{(E)}$ to be the state of the ego vehicle in the ground truth data at time $t$, and define $\mathbf{\wh{x}}_t^{(E)}$ to be the state of the ego vehicle controlled by the planner, at time $t$. For any given scene, we compute
    \begin{equation*}
        \frac{\min\limits_t \min\limits_{i \in [N]} \norm{\mathbf{\wh{x}}_t^{(E)} - \mathbf{x}_t^{(i)}}_{xy}}{\min\limits_t \min\limits_{i \in [N]} \norm{\mathbf{x}_t^{(E)} - \mathbf{x}_t^{(i)}}_{xy}},
    \end{equation*}
    where $\norm{\cdot}_{xy}$ computes the norm only along the $x$ and $y$ spatial coordinates of the given state vector. We average this quantity across all scenes. Intuitively, this is a measure of how close the planner is willing to get to other vehicles, compared to the ground truth ego vehicle movement.

    \item \underline{Progress:} Let $\mathbf{x}_g$ denote the goal state of the ego vehicle, and let $\mathbf{x}_s$ denote its starting state. Next, let $\mathbf{x}_f$ denote the final state of the ego vehicle once the planner no longer provides any new plans. We define progress as
    \begin{equation*}
        1 - \frac{\norm{\mathbf{x}_g - \mathbf{x}_f}_{xy}}{\norm{\mathbf{x}_g - \mathbf{x}_s}_{xy}}.
    \end{equation*}
\end{itemize}

\subsection{Un-Calibrated Confidence Set Metrics, Waymo Dataset}
Below, we compare the coverage rate and set sizes between the un-calibrated version of our algorithm, using only quantile regression as a confidence-set-generating tool, and the calibrated version of our algorithm, leveraging conformal prediction.
\begin{table}[H]
  \centering
  \scriptsize
  \caption{Un-Calibrated Confidence Set Coverage Metrics}
    \label{tab:uncalibrated_set_size}
  \begin{tabular}{lab}
    \toprule
    \textbf{Prediction Step} & \textbf{Calibrated Set} & \textbf{Un-Calibrated Set} \\
    & \textbf{Coverage Rate} & \textbf{Coverage Rate} \\
    \midrule
    0.5 & 0.997 \scriptsize{$\pm 0.002$} & 0.968 \scriptsize{$\pm 0.007$} \\
    1.0 & 0.986 \scriptsize{$\pm 0.006$} & 0.887 \scriptsize{$\pm 0.015$} \\
    1.5 & 0.980 \scriptsize{$\pm 0.008$} & 0.839 \scriptsize{$\pm 0.019$} \\
    2.0 & 0.967 \scriptsize{$\pm 0.011$} & 0.817 \scriptsize{$\pm 0.022$} \\
    2.5 & 0.965 \scriptsize{$\pm 0.011$} & 0.818 \scriptsize{$\pm 0.025$} \\
    3.0 & 0.965 \scriptsize{$\pm 0.013$} & 0.800 \scriptsize{$\pm 0.028$} \\
    \bottomrule
  \end{tabular}
\end{table}

\begin{table}[H]
  \centering
  \scriptsize
  \caption{Un-Calibrated Confidence Set Sizes}
    \label{tab:uncalibrated_set_size}
  \begin{tabular}{lab}
    \toprule
    \textbf{Prediction Step} & \textbf{Calibrated} & \textbf{Un-Calibrated} \\
    & \textbf{Set Size} & \textbf{Set Size} \\
    \midrule
    0.5 & 61 \scriptsize{$\pm 6$} & 42 \scriptsize{$\pm 4$} \\
    1.0 & 246 \scriptsize{$\pm 27$} & 153 \scriptsize{$\pm 16$} \\
    1.5 & 655 \scriptsize{$\pm 71$} & 406 \scriptsize{$\pm 46$} \\
    2.0 & 1361 \scriptsize{$\pm 143$} & 815 \scriptsize{$\pm 86$} \\
    2.5 & 2422 \scriptsize{$\pm 240$} & 1434 \scriptsize{$\pm 145$} \\
    3.0 & 3885 \scriptsize{$\pm 365$} & 2300 \scriptsize{$\pm 226$} \\
    \bottomrule
  \end{tabular}
\end{table}

We primarily notice that conformal prediction enforces that set coverage remains above the desired 95\% threshold, although this comes at the cost of providing larger sets. Additionally, conformal prediction aids in reducing the variance of coverage rates, especially for later-timestep predictions. This demonstrates the $\cO\rp{1/T}$ convergence guarantee on the deviation of the realized, empirical error rate from the desired error rate -- with more data, the absolute deviation in conformal prediction's error rate from the desired error rate decreases, whereas this is not guaranteed for quantile regression alone.

\subsection{Planner Visualization}
\begin{figure}[H]
    \vspace{0.0625in}
    \centering
    \includegraphics[scale=0.7]{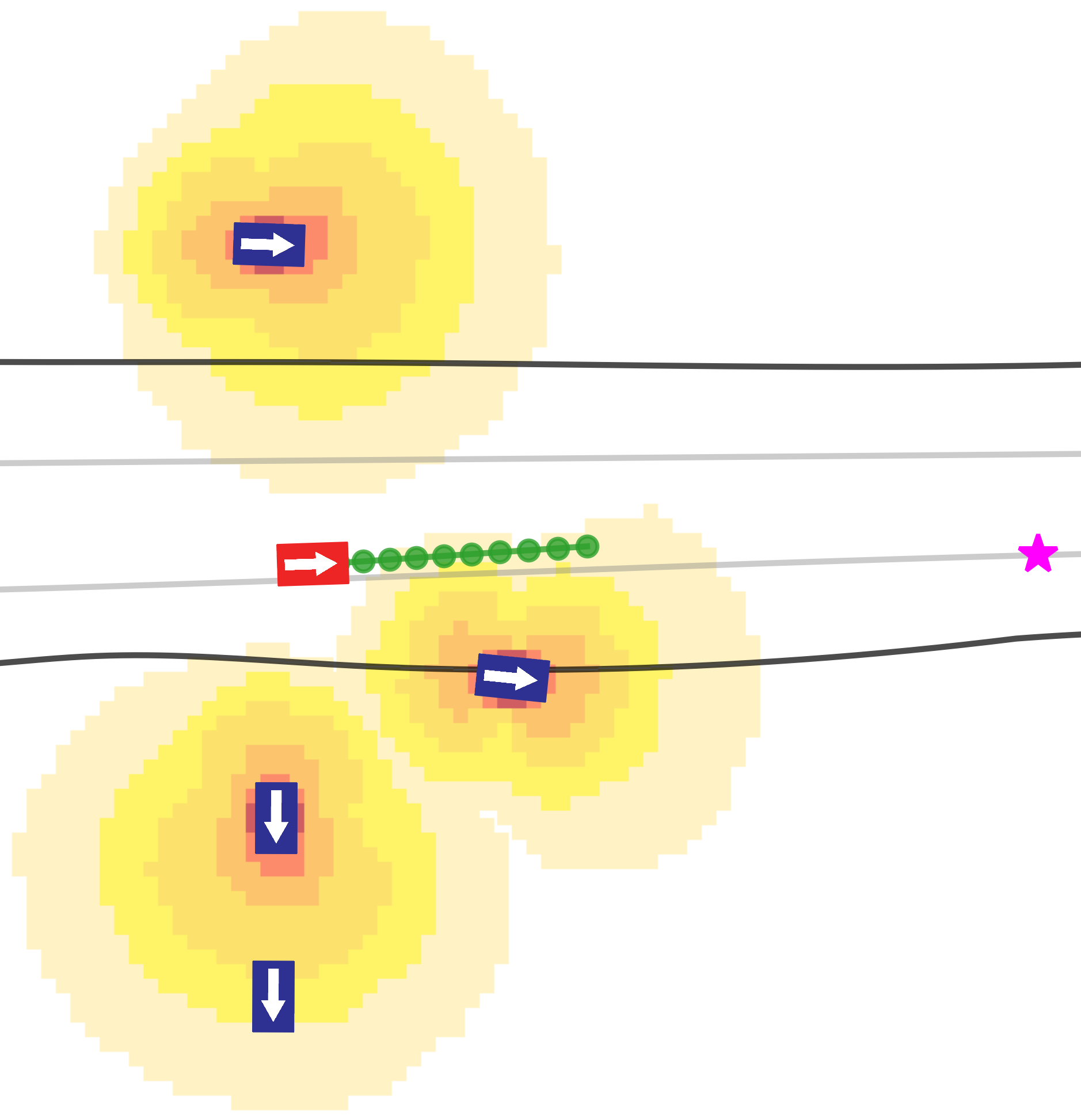}
    \caption{Visualization of the planner. The autonomous ego vehicle is shown in red, and the human drivers are shown in blue. The plan is shown in green, representing the tracking of the original HJ-generated plan using iLQR.}
\end{figure}

\subsection{Additional Algorithm Implementation Details}
We opt for the version of Trajectron++ without the encoder for maps and the encoder for future ego-agent motion plans due to the lack of availability of these in the datasets. The model architecture and hyperparameters are kept the same as in \cite{salzmann2020trajectron++}.

We explored graph-based planning algorithms such as A* and Dijkstra but found them to be computationally intractable as a result of the high dimensionality of the Extended Dubins' car model and the presence of time-varying dynamic obstacles. Our reachability-based planner does not suffer from such issues.

In the case in which no feasible plan to the desired target state exists, for example due to the large sizes of the probabilistic reachable sets, our reachability-based planner produces a plan minimizing the distance between the final ego state and the desired target state.

\end{document}